# Electrical detection of high-order optical orbital angular momentum

Guanyu Zhang[1], Xianghan Meng[1], Zini Cao[1], Hai Lin[2], Shuxin Huang[1], Minghao Deng[1], Jiaqi Li[1], Qihuang Gong[1, 3, 4], and Guowei Lyu[1, 3, 4*]

*[1] State Key Laboratory for Mesoscopic Physics, Collaborative Innovation Center of Quantum Matter, Frontiers Science Center for Nano-Optoelectronics, School of Physics, Peking University, Beijing 100871, China*

*[2] School of energy and physics, Huizhou University, Huizhou, Guangdong 516007, China*

*[3] Collaborative Innovation Center of Extreme Optics, Shanxi University, Taiyuan, Shanxi 030006, China*

*[4] Peking University Yangtze Delta Institute of Optoelectronics, Nantong, Jiangsu 226010, China*

** Corresponding author: guowei.lyu@pku.edu.cn*

## Abstract

The orbital angular momentum (OAM) of light provides an unbounded set of orthogonal modes for ultrahigh-capacity optical information processing. However, current OAM detection schemes typically rely on light interference or diffraction, which require bulky optical components and pose a major obstacle to on-chip integration. Here, we demonstrate a fully integrated silicon-based photodetector that enables direct electrical detection of light OAM. This photodetector can resolve vortex beams with topological charges from m = −9 to 9, achieving a record-high mode number resolution among on-chip devices. By integrating plasmonic gratings onto the device electrodes, incident vortex beams can be converted into surface plasmon polaritons with OAM-dependent splitting angles, which in turn produce photocurrents that vary monotonically with the OAM order. Further incorporation of a surface dielectric lens can enhance mode resolution, and a split-electrode architecture enables OAM chirality discrimination. Owing to its CMOS-compatibility and spectral scalability, this platform provides a compact and robust solution for integrated OAM detection, opening new opportunities for on-chip optical communication and computing systems based on structured light.

## Introduction

Optical orbital angular momentum (OAM), carried by vortex beams with a helical phase structure, constitutes an independent degree of freedom of light and has attracted growing attention for its wide-ranging applications in optomechanical manipulation[1,2], super-resolution imaging[3], sensing[4,5], and quantum information processing[6]. OAM theoretically supports an infinite set of mutually orthogonal optical modes, offering great potential to enhance the capacity and security of both classical[7,8] and quantum[9] optical communications. Harnessing these advantages, however, critically depends on the availability of compact, high-speed, and multi-mode OAM detection technologies capable of operating in integrated photonic platforms.

Conventional OAM detection methods mostly rely on light interference[10] and diffraction[11], where the OAM state is determined by analyzing interference fringes or diffraction patterns. Alternatively, spatial light modulators combined with Fourier optics can convert OAM states into distinct spatial positions for identification[12]. However, these approaches typically require bulky optical setups, limiting their compatibility with on-chip integration. Advances in nanophotonics have enabled metasurfaces as compact alternatives to spatial light modulators[13,14], yet most metasurface-based schemes still depend on imaging systems for OAM readout. Surface plasmon polaritons (SPPs) offer a promising route toward miniaturized OAM detection, as vortex-beam excitation can impart OAM to SPPs[15]. Properly engineered nanostructures, such as annular gratings[16,17] or apertures[18], can spatially separate SPPs associated with different OAM states. Nevertheless, these techniques continue to rely on far-field imaging[19] or even near-field probing[20] systems. Hence, integrated OAM photodetectors with multi-mode detection capability and direct electrical readout remain highly sought after.

Although the optical multiplexing and discrimination of vortex beams have been extensively explored, direct electrical detection of light OAM remains scarce. Existing approaches can be broadly classified into SPP-based[21] and orbital photogalvanic effect (OPGE)-based photodetectors. In SPP-based schemes, a spin–Hall plasmonic grating was integrated with a $PdSe_2$-based photothermal detector acting as a spot position sensor[22], enabling the conversion of spatially separated SPP foci into electrical signals and achieving OAM detection up to |m| = 4. On the other hand, OPGE-based detectors exploit a second-order photoelectric effect driven by the helical phase gradient of light, generating photocurrents proportional to the OAM order. This mechanism was first observed in low-symmetry two-dimensional $WTe_2$[23], and later reported in $TaIrTe_4$[24] and multilayer graphene[25,26]. While OPGE provides a direct link between OAM and photocurrent, it typically requires differential measurements under left- and right-circularly polarized illumination, making the method noise-sensitive and unsuitable for high-speed communication. To date, reported on-chip OAM detection schemes are limited to resolving only a few low-order modes ($|m| \leq 4$) and often rely on novel two-dimensional materials, for which large-scale fabrication technologies are still under development.

In this work, we present a silicon-based photodetector that directly converts optical OAM into electrical signals using a compact grating architecture. The device is capable of resolving vortex beams with OAM orders up to |m| = 9, achieving a record-high mode number resolution for on-chip OAM detection. Incident vortex beams with different OAM orders are coupled into SPP modes propagating along distinct directions and experiencing varying losses, which are subsequently transduced into OAM-dependent photocurrents at a Schottky junction. Incorporating a surface dielectric lens further enhances the discrimination between adjacent OAM states. Additionally, a split-electrode configuration enables the determination of the OAM chirality by comparing the relative intensities of the SPP branches. The photodetector exhibits rise and fall times of 1.7 μs and 1.83 μs, respectively, supporting high-speed electrical readout without polarization modulation. The device achieves an average OAM responsivity of 226 $nA \cdot W^{-1}$ and exhibits a noise-equivalent OAM resolution down to 0.05 $Hz^{-1/2}$, outperforming previously reported studies. These results establish a robust, CMOS-compatible platform for integrated OAM detection and open new opportunities for high-capacity optical communication, sensing, and information processing based on structured light.

# Results

## On-chip OAM photodetector as a proof of concept

The orbital angular momentum of light arises from the helical phase front that winds around the beam axis and is characterized by the quantized OAM order, i.e., topological charge $m$. It can be expressed as the phase gradient integral looping around the beam's central singularity, $m = \frac{1}{2\pi}\oint_C \varphi(\boldsymbol{r})dr$, where $\varphi$ is the optical phase, and $r$ is the position vector. For a vortex beam, the helical phase front implies that the local wave vector is tilted with respect to the propagation direction, yielding an in-plane component, $k_{OAM} = k \cdot \frac{n_{\parallel}}{|\boldsymbol{n}|}$ , where $k$ is the magnitude of the free-space wave vector, $\boldsymbol{n}$ is the unit normal to the wavefront, and $n_{\parallel}$ is its projection parallel to the detector plane. As $m$ increases, the phase winds more rapidly within one wavelength, leading to stronger front twisting and consequently a larger in-plane wave vector (see Supplementary Note 1 for details).

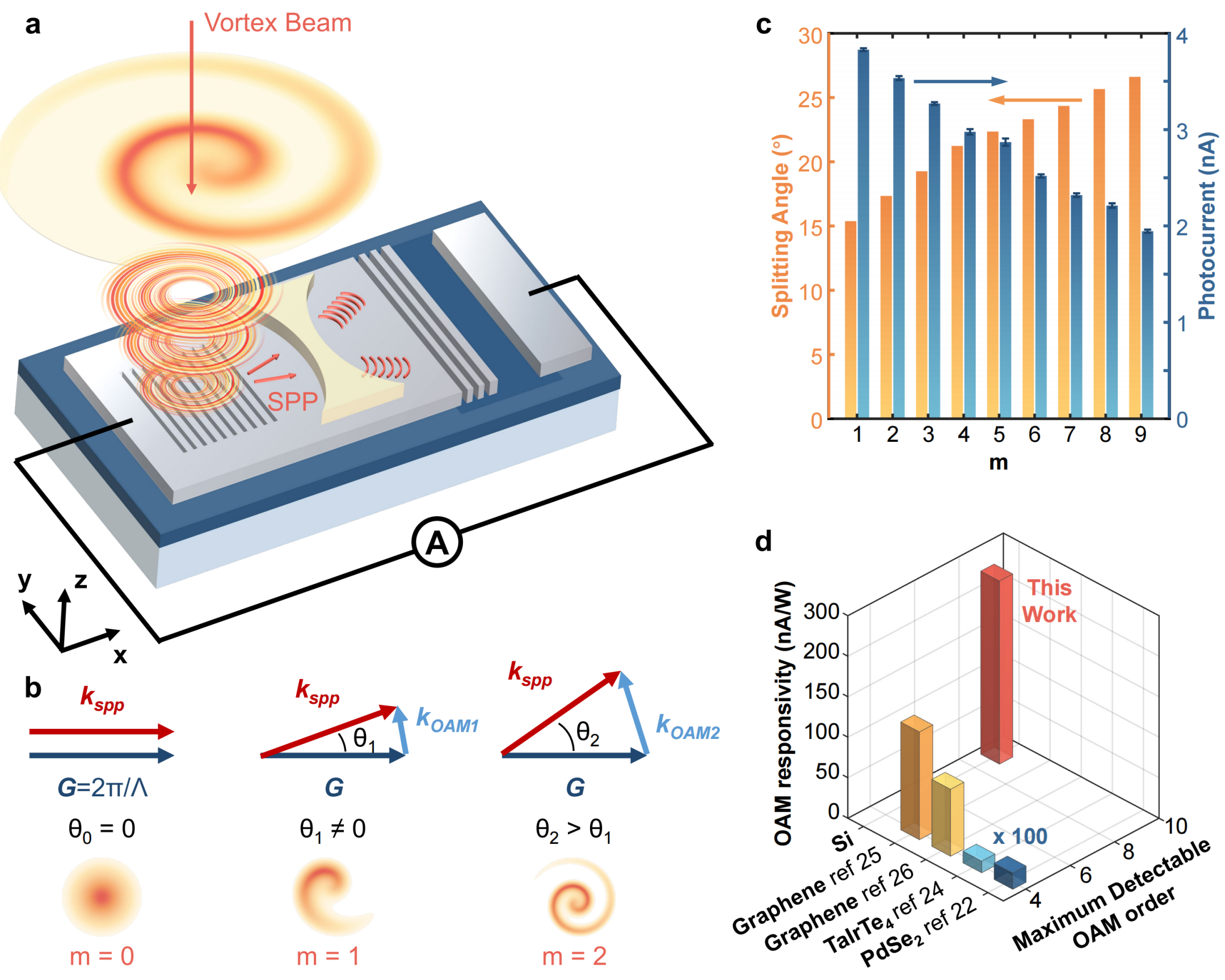

**Fig. 1| Device architecture and operational principle of the OAM photodetector. a.** Schematic illustration of the OAM photodetection methods. The incident vortex beam couples to specific SPPs via the plasmonic grating, generating an encoded photocurrent response. **b.** Vector diagrams representing the momentum matching condition for vortex beams with $m$ = 0, 1, and 2. The SPP propagation angle increases with the OAM order. **c.** Simulated splitting angles (orange) and measured photocurrents (blue) for incident vortex beams with OAM orders $m$ ranging from 1 to 9. Error bars indicate standard deviations. **d.** Comparison of OAM responsivity and maximum detectable OAM order against state-of-the-art direct OAM photodetectors.

Fig. 1a schematically illustrates the concept of the on-chip OAM photodetector, which employs a planar silicon-based architecture incorporating input and output coupling gratings on one side of the electrode. Both gratings are one-dimensional linear structures, and a dielectric lens is integrated between them to expand the detectable OAM range. During measurement, a slightly focused vortex beam is normally incident at the center of the input grating, exciting SPPs that satisfy the momentum-matching condition (Fig. 1b), $\boldsymbol{k}_{\mathbf{spp}} = \boldsymbol{G} + \boldsymbol{k}_{\mathbf{OAM}}$, where $\boldsymbol{G}$ is the reciprocal lattice vector of the grating and $\boldsymbol{k}_{\mathbf{spp}}$ is the plasmon wave vector. In the experiments, four positions on the vortex beam fulfilled this condition, resulting in four SPP branches propagating in distinct directions (Supplementary Fig. S1). Two of them, traveling toward the electrode–silicon interface, contribute to the photocurrent, separated by splitting angles ±θ relative to the x-axis. As $m$ increases, the in-plane component $\boldsymbol{k}_{\mathbf{OAM}}$ increases, leading to a larger θ according to the law of cosines, as confirmed by simulation (Fig. 1c). Consequently, SPPs excited by higher-order OAM beams propagate over longer distances and undergo greater loss. These SPPs are subsequently reconverted into optical fields by the output grating at the electrode–silicon interface and then transformed into photocurrents via the photovoltaic effect at the Schottky junction. The photocurrent amplitude decreases with increasing $m$ (Fig. 1c). By exploiting this OAM-resolved plasmonic coupling mechanism, the device discriminates vortex beams from $|m|$ = 1 to 9 with an average OAM responsivity of $R_{OAM} = \frac{\Delta I_{ph}}{P \cdot \Delta m}$, where $I_{ph}$ is the photocurrent, and $P$ is the incident power. The measured $R_{\mathrm{OAM}}$ reaches 226 nA $W^{-1}$, surpassing all previously reported results (Fig. 1d).

To experimentally validate the OAM-resolving grating scheme, we first designed and fabricated a baseline photodetector as shown in Fig. 2. The input grating period Λ was selected to satisfy the coupling condition as $|\boldsymbol{k}_{\boldsymbol{SPP}}| - |\boldsymbol{k}_{\boldsymbol{OAM}}| < \frac{2\pi}{\Lambda} < |\boldsymbol{k}_{\boldsymbol{SPP}}| + |\boldsymbol{k}_{\boldsymbol{OAM}}|$. In subsequent experiments, Λ = 610 nm was chosen such that $\boldsymbol{G} \approx \boldsymbol{k}_{\boldsymbol{OAM}}$ at the 633 nm excitation wavelength. Under this condition, the SPP splitting angle θ increases from zero as the incident OAM order $m$ increases. The grating parameters were optimized through finite-difference time-domain (FDTD) simulations. Maximum coupling efficiency was achieved at a slot width of 80 nm and a depth of 120 nm (Supplementary Fig. S2). The input grating length $l_{gra}$ was set to 12 μm with 20 periods, ensuring full coverage of the vortex beam (Fig. 2c). To balance responsivity and OAM resolution, the center-to-electrode distance $D$ was designed as 30 μm, while the electrode width $w_{ele}$ was set to 40 μm to suppress edge reflection and scattering. The output grating shared the same period as the input grating and was etched along the electrode edge.

FDTD simulations of the intensity on the electrode surface under vortex illumination with $m = +1$ to +9 are shown in Fig. 2a (Supplementary Note 2). The averaged intensity profiles along the electrode edge, normalized to the peak value at $m = 1$, are presented in the side panel. Here, a vortex beam with clockwise rotation is defined as positive. The excited SPPs form two counter-propagating branches directed upward and downward. With increasing $m$, the separation between the branches widens progressively but with a diminishing rate, which is consistent with the gradual saturation of $\boldsymbol{k}_{OAM}$. It is worth noting that the angular distribution of the SPP emission spans a broad range. This arises from spatial variations in the vortex wavefront, which actually contains a continuum of $\boldsymbol{k}_{OAM}$ components. The SPP splitting angle θ was extracted along the direction of maximum intensity, and theoretical calculations were performed using the same criterion (Supplementary Note 1). The simulated and theoretical results (Fig. 2e) show excellent agreement, confirming the OAM discrimination mechanism. As $m$ increases, the total SPP power reaching the electrode edge decreases due to enhanced propagation loss. Additionally, the rate of this decrease diminishes at higher $m$, which is consistent with the expectations. Moreover, the downward-propagating branch exhibits notably higher intensity than the upward branch, attributed to their distinct excitation positions. SPPs in the upward branch are launched further from the electrode–silicon interface and experience stronger grating losses. When the chirality of the incident vortex beam is reversed, the relative intensities of the two SPP branches also invert.

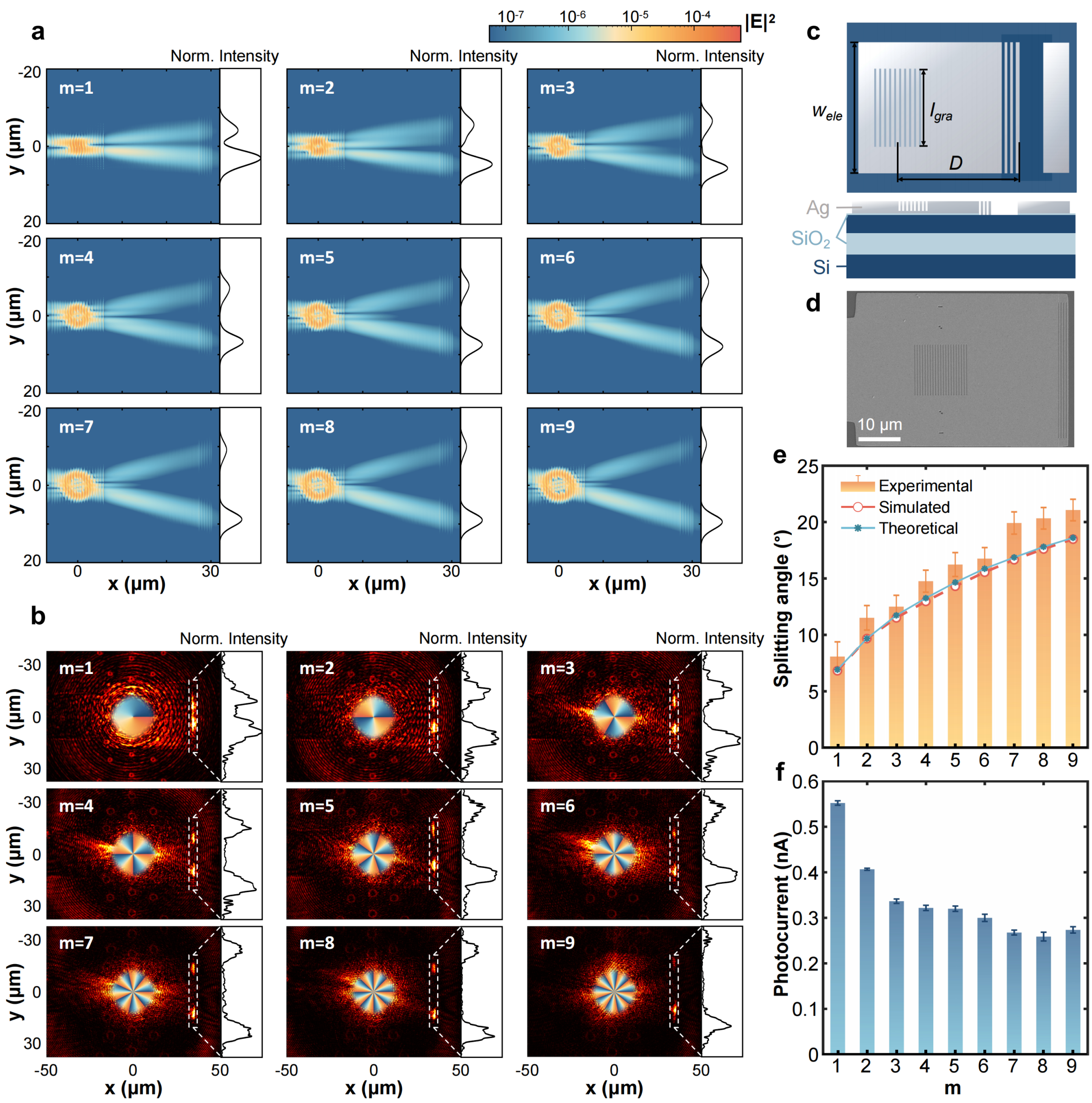

**Fig. 2| Numerical simulation and experimental characterization of the grating-based OAM photodetector. a.** Simulated light intensity on the silver electrode surface under vortex beams with $m$ =1-9. Side panels: corresponding normalized intensity profiles along the $y$-axis at the output grating position. **b.** Experimentally measured intensity distributions for $m$ = 1 to 9. The white dashed boxes denote the output coupling grating region where SPP splitting is observed. Side panels: integrated intensity profiles within the boxed regions, normalized to the peak intensity of the $m$ = 1

mode. **c.** Top-view (upper) and cross-sectional (lower) schematics of the device architecture. **d.** Scanning electron microscopy (SEM) image of a representative fabricated device. **e.** Comparison of the SPP splitting angles obtained from theoretical calculations (blue), numerical simulations (red), and experimental measurements (orange bars) for *m* = 1-9. Error bars represent the average full width at half maximum of measurements. **f.** Measured photocurrent as a function of the incident OAM order *m*. Error bars indicate standard deviations.

Devices with the optimized parameters were fabricated on a silicon-on-insulator (SOI) substrate. A $SiO_2$ insulating layer was first deposited, with the region between the two electrodes selectively etched to form a Schottky contact. The insulating layer prevents short-circuiting through the top silicon layer in non-active areas, and also prevents the possible excessive etching of input gratings during fabrication. The electrodes were patterned using electron-beam lithography followed by electron-beam evaporation, while the gratings were defined by focused ion beam milling (see Methods for details). To avoid spurious photocurrent generation, the input grating was designed not to penetrate the metal electrode (Supplementary Fig. S3). In contrast, the output grating was etched through the electrode to enhance optical-to-electrical conversion efficiency. An SEM image of a representative device is shown in Fig. 2d. During measurements, vortex beams were generated using a Q-plate. After being slightly focused by an objective lens, the beam was normally incident on the input grating, exciting SPPs that propagated toward the electrode-silicon interface. A portion of the SPP field was outcoupled into free space via the output grating and collected along the optical path for imaging (Supplementary Fig. S4).

The measured SPP mode distributions under incident beams with *m* = 1–9 are presented in Fig. 2b.

The side panel shows the average normalized light intensity along the $y$-direction at the electrode edge, corresponding to the white box in the image. The experimental results agree closely with both theoretical and simulated predictions: as the OAM order $m$ increases, the SPP beam splitting angle enlarges, while the edge intensity gradually decreases. Minor fluctuations at the intensity peaks are attributed to stripe noise caused by reflected light from the grating captured by the CMOS. At higher $m$, the splitting angle exhibits a saturation trend, consistent with theoretical expectations (Fig. 2e). The photoresponse of the device is shown in Fig. 2f. As $m$ increases, the photocurrent gradually decreases, demonstrating direct OAM-to-electrical conversion. However, as revealed by the measurements (Fig. 2f) and the simulations (Fig. 3a), the photocurrent variation rate rapidly converges toward zero with increasing $m$. Consequently, for vortex beams with $m > 4$, the photocurrent change induced by a one-unit difference in OAM order becomes comparable to the noise level, hindering effective discrimination. Additional structural optimization is therefore required to enhance the distinction between SPP modes excited by higher-order vortices and to further expand the detectable OAM range.

## Enhancing the detection capacity through dielectric lenses

Depositing a dielectric film on a metal surface modifies the propagation constant of SPPs, changing their effective refractive index. Consequently, patterned dielectric layers can serve as plasmonic lenses to focus or diverge SPP waves. Their focal length follows an expression analogous to that of traditional lenses: $\frac{1}{f} = (\Omega - 1)\left(\frac{1}{R_1} - \frac{1}{R_2} + \frac{(\Omega-1)t}{n_2 R_1 R_2}\right)$, where $f$ is the plasmonic focal length, $R_1$ and $R_2$ are the radii of curvature on both sides, $n_1$ and $n_2$ denote the real parts of the effective indices in regions without and with the dielectric, $t$ is the central thickness, and $\Omega = n_2/n_1$. Introducing a concave plasmonic lens between the input and output gratings (Fig. 3a) enlarges the SPP splitting angle and increases the propagation-loss contrast, thereby enhancing the discrimination of high-order OAM states (Supplementary Fig. S5). Geometrically, SPPs excited by high-order vortices also propagate longer distances inside the lens, experiencing stronger attenuation. Collectively, these effects endow the lens-integrated device with a markedly improved resolving capability when compared to the baseline design, as evidenced in the normalized transmission efficiencies shown in Fig. 3b.

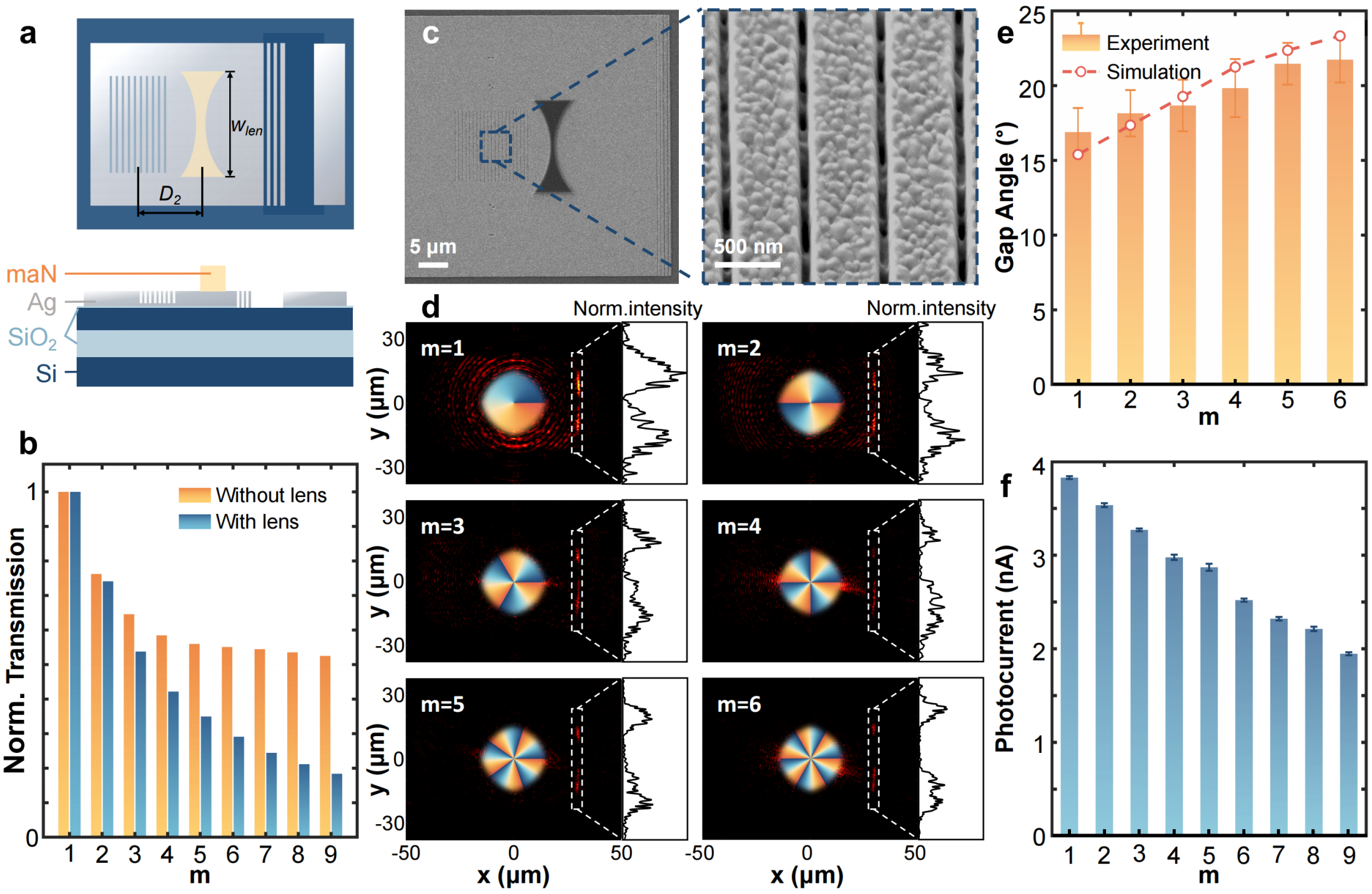

**Fig. 3| Optical and optoelectronic characterization of the lens-integrated OAM photodetector.** **a.** Top-down (upper) and cross-sectional (lower) schematic of the photodetector incorporating a plasmonic lens **b.** Simulated normalized transmission as a function of OAM order *m* for devices with (blue) and without (orange) the integrated lens. **c.** SEM images of the fabricated lens-integrated

device. The inset provides a magnified view of the input coupling grating structures. **d.** Measured optical intensity distributions for OAM orders *m* = 1-6. The contrast of the images has been enhanced for visualization. The white dashed boxes indicate the output grating region. Side panels: corresponding normalized intensity profiles along the *y*-axis, showing the spatial separation of the SPPs. **e.** Simulated (red) and experimentally measured (orange) SPP splitting angles for *m* = 1-6. Error bars represent the average full width at half maximum of the measured SPP branches. **f.** Measured photocurrent response of the optimized photodetector under vortex beam illumination for OAM orders ranging from 1 to 9. Error bars indicate standard deviations.

Experimentally, an additional electron-beam exposure step was used to pattern a concave dielectric lens made of maN photoresist, as shown in the SEM image in Fig. 3c. The optimized focal length f=8 μm and the center-to-grating distance $D_2$=10 μm were obtained through global FDTD optimization. Following the same optical measurement procedure, the SPP splitting angle and photocurrent of the new device were characterized. As shown in Fig. 3d and e, the SPP splitting angle increases significantly, in agreement with the simulations. Here, the splitting angle is also defined as the angle between the line connecting the output SPP lobe and the vortex-beam center and the x-axis for consistency. Due to reflection and interference at the lens interface, each SPP branch becomes broadened and exhibits multi-peak intensity features, consistent with the simulation results in Supplementary Fig. S6. The measured photocurrent (Fig. 4f) decreases steadily with increasing OAM order and follows an overall linear trend, enabling direct electrical readout of vortex beams from *m* = 1 to *m* = 9. The average OAM responsivity reaches $R_{OAM}$=226 nA·$W^{-1}$, significantly exceeding previously reported values. The device also exhibits an exceptionally low dark current of 0.95 pA, yielding a noise-equivalent OAM resolution as low as NEΔm= $I_{noise}/(\mathrm{d}I_{ph}/\mathrm{d}m)$ = 0.05 $Hz^{-1/2}$ at the -3dB cut-off frequency (noise spectrum in Supplementary Fig. S7).

Compared with commercial silicon detectors, the response speed is much slower, with rise and fall times of 605 μs and 845 μs, respectively (Supplementary Fig. S8), corresponding to a cutoff frequency of ~572 Hz. Further measurements indicate that the reduced speed is attributable to parasitic capacitance induced by the 20 nm $SiO_2$ insulating layer between the electrode pad and the SOI substrate. Removing the top-silicon layer in non-active regions not only prevents electrical shorting but also suppresses this parasitic capacitance. Based on this approach, a high-speed OAM photodetector was fabricated (Supplementary Fig. S9) and achieved a rise/fall time of 1.70/1.83 μs, corresponding to approximately 198 kHz bandwidth (Supplementary Fig. S9). As this value approaches the modulation limit of the laser source used, the intrinsic device speed may be even higher. However, the responsivity of this high-speed device is substantially decreased, which limits its applicability in direct detection of high-order OAM modes (Supplementary Fig. S10). With a refined electrical design, we anticipate that both parasitic capacitance and excess noise can be effectively suppressed, enabling a high-speed OAM photodetector for robust vortex-beam sensing.

## Chirality Discernibility in a Split-Electrode Device

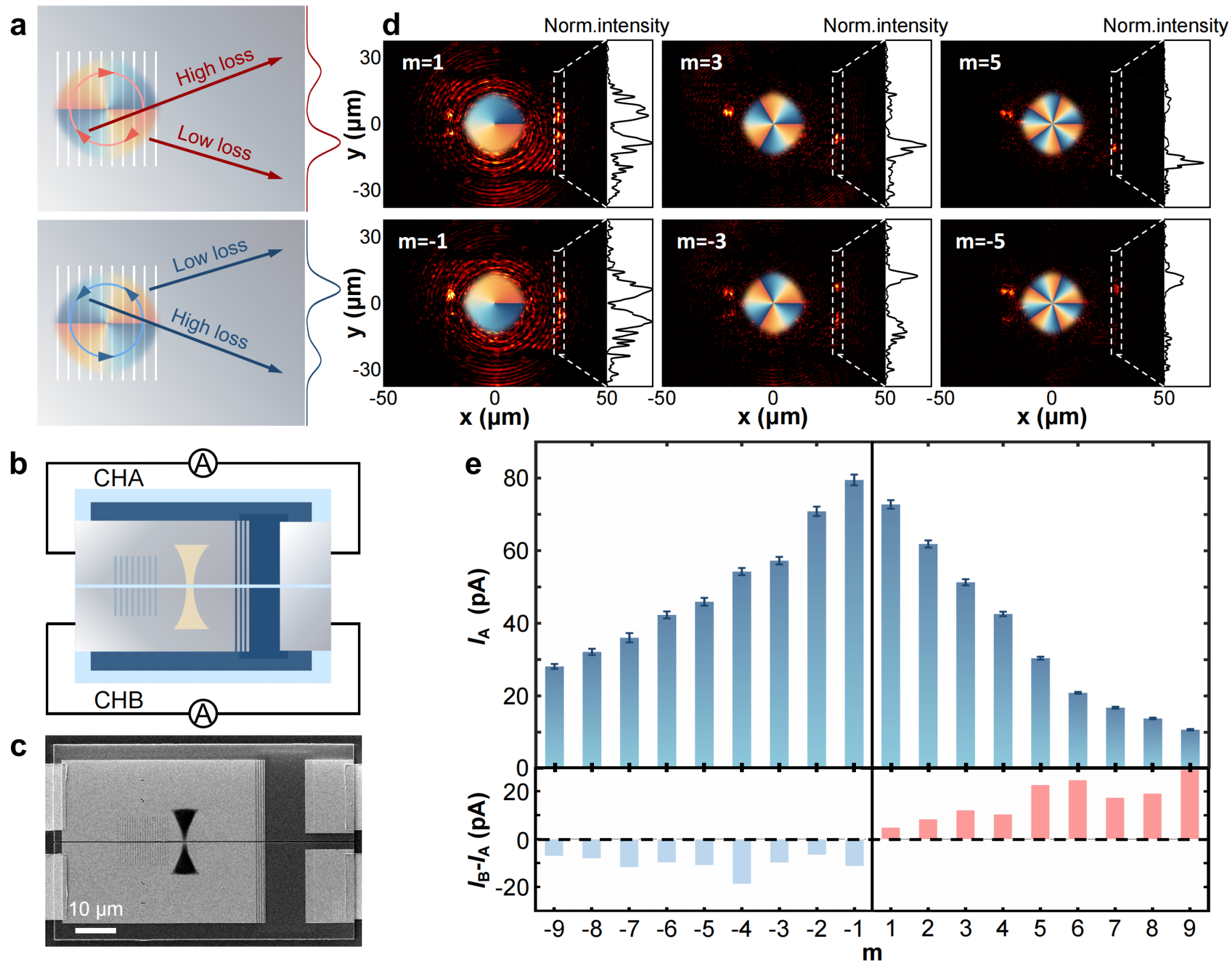

**Fig. 4| Discrimination of OAM chirality in a split-electrode photodetector. a.** Operational principle for chirality-resolved OAM detection. Asymmetric loss during SPP propagation enables the separation of vortex modes with opposite chirality. **b.** Device layout of the split-electrode photodetector defining channel A (CHA) and channel B (CHB). **c.** SEM image of the fabricated split-electrode photodetector. **d.** Intensity distributions of the device under vortex beams with opposite chirality ($m = \pm1, \pm3, \pm5$). The contrast has been enhanced for visualization. White dashed boxes indicate the output grating regions. Side panels: corresponding normalized intensity profiles along the $y$-axis. **e.** Photoresponse of the split-electrode device. Upper panel: measured photocurrent in Channel A ($I_A$) as a function of OAM order $m$ from -9 to +9. Lower panel: photocurrent difference between the two channels ($I_B$ - $I_A$). Error bars represent standard deviations.

Reversing the chirality of the incident vortex beam inverts the relative intensities of the excited SPP branches while maintaining the total SPP power that reaches the Schottky junction. Consequently, the single-electrode configuration discussed above can only resolve the magnitude of the OAM order ($|m|$), with the sign remaining indeterminate. To overcome this limitation, a modified device architecture was proposed to distinguish the spatial intensity asymmetry of the SPP branches, thereby enabling unambiguous chirality detection. The working principle is illustrated in Fig. 4a. For a vortex beam with positive OAM order $+m$, momentum matching excites two SPP branches propagating toward the electrode–silicon interface. Due to geometric asymmetry, the upward-propagating branch traverses more grating periods and suffers higher propagation loss than the downward branch. This intensity disparity, confirmed by previous simulations and experiments (Figs. 2b and 3d), inverts when the OAM chirality is reversed (Supplementary Fig. S11). Therefore, monitoring the intensity contrast between the two branches allows for chirality discrimination.

We implemented this concept using a split-electrode configuration (Fig. 4b), where the top silicon layer, Ag electrode, grating, and lens are physically bisected into two electrically isolated sections. These upper and lower halves function as independent photodetectors (Supplementary Fig. S12), connected to channels A and B of a picoammeter for simultaneous measurement ($I_A$ and $I_B$). Fig. 4c shows an SEM image of the fabricated device. To clearly visualize the SPP distribution, Fig. 4d presents the extracted intensities for a reference device without a dielectric lens. For $+m$ inputs, the downward SPP branch is significantly more intense than the upward one, with the contrast becoming more pronounced at higher $|m|$. For $-m$, this distribution is reversed. The corresponding photocurrent response (Fig. 4e) confirms the behavior of $I_A > I_B$ for positive OAM and $I_B > I_A$ for negative OAM. Simultaneously, the overall photocurrent magnitude decreases with $|m|$, preserving the capability to determine the OAM order. Thus, the device successfully resolves both OAM chirality and magnitude from $m$ = -9 to +9. Although our measurements were limited to $|m| \leq 9$ by the generation limit of the Q-plate, the device is theoretically capable of resolving higher-order OAM modes, provided the photocurrent variation induced by OAM changes exceeds the system noise floor.

## Discussion

In summary, we have demonstrated a compact silicon-based photodetector that directly converts the optical OAM into electrical signals. By encoding the OAM order of vortex beams into the splitting angle and propagation loss of SPPs, the device can reliably discriminate OAM states ranging from m = -9 to +9, achieving the record-high distinguishable mode numbers reported for on-chip OAM detection. Numerical simulations indicate the potential to expand the detectable OAM range to higher orders. Meanwhile, the improvement of the silver coating and the optimization of the output gratings are expected to enhance the OAM responsivity further. The current device bandwidth reaches 198 kHz at the cost of signal-to-noise ratio, and the response speed is primarily limited by parasitic capacitance. Further optimization of the electrical layout is expected to alleviate these limitations, opening a pathway toward MHz-bandwidth operation.

Beyond performance improvements, this detection scheme offers substantial flexibility in spectral expansion and system-level integration. The grating-mediated coupling mechanism is inherently tunable, which enables a straightforward adaptation to telecommunication wavelengths through geometric rescaling and the incorporation of suitable narrowband semiconductors. Combined with its CMOS compatibility and direct electrical readout, this platform is well-suited for seamless integration with electronic and photonic circuits, promising its applications in high-dimensional quantum information processing[27], on-chip optical neural networks[28], high-capacity optical communication[4], and multidimensional imaging[29], et al. More comprehensively, this work lays a scalable foundation for leveraging the unlimited degree of freedom provided by optical OAM in integrated photonic technologies.

# Methods

## Device Fabrication

The chirality-resolved devices (Fig. 4) were fabricated on SOI substrates with a 500-nm n-type top silicon layer. First, the split-silicon structures were defined by electron-beam lithography (EBL; ELS-F125, Elionix, Japan) using maN-2405 resist, followed by inductively coupled plasma (ICP) etching (PlasmaPro 100 Cobra, Oxford Instruments, England) down to the buried oxide. A 20-nm $SiO_2$ insulating layer was deposited via atomic layer deposition (ALD; MNT-S200Oz, MNT Micro and Nanotech, China). To expose the active silicon contact areas, a second EBL step (950 PMMA) and subsequent $SiO_2$ ICP etching were performed. The plasmonic excitation and propagation regions were defined by a third EBL step (950 PMMA), followed by electron-beam evaporation of Ti/Ag (5/160 nm) and lift-off. Peripheral electrodes were patterned via a fourth EBL process, followed by magnetron sputtering of Ti/Ag (5/50 nm) and lift-off. Input and output coupling gratings were milled using a focused ion beam (FIB; Crossbeam 540, Zeiss, Germany). A 3-nm $Al_2O_3$ passivation layer was deposited by ALD to prevent silver oxidation. Finally, the dielectric lenses were patterned using EBL with maN-2405 resist.

Fabrication for the lens-integrated devices (Fig. 3) followed a simplified protocol: the initial silicon etching step was omitted, and the plasmonic region and peripheral electrodes were patterned in a single EBL/evaporation step (Ti/Ag, 5/160 nm), eliminating the need for an additional EBL and sputtering. For the baseline devices (Fig. 2), the process was further streamlined by excluding the final dielectric lens patterning.

### Optoelectronic Characterization

Measurements were performed using a custom-built optoelectronic setup. A continuous-wave 633 nm HeNe laser (16194, REO, USA) beam was directed through an optical switch and a quarter-wave plate, then converted into vortex beams with distinct OAM orders using a Q-plate array (VRA-633, LBTEK, China). OAM orders were verified using an interferometric setup, with interference patterns recorded by a CMOS camera (Supplementary Fig. S13). The vortex retarder array was mounted on a motorized XY stage to allow rapid switching between OAM states. A linear polarizer aligned the beam polarization parallel to the grating vector for optimal coupling, and the beam was focused normally onto the input grating using an objective lens.

Photocurrents were recorded with a dual-channel picoammeter (6482, Keithley, USA). Noise spectra and temporal responses were characterized using a lock-in amplifier (MFLI, Zurich Instruments, Switzerland) and an oscilloscope (4262, Pico Technology, England), respectively. Incident power was calibrated to around 1 mW using a power meter (PM101, Thorlabs, USA). SPP propagation images were captured using a CMOS camera.

## Data availability

All data needed to evaluate the conclusions in the paper are present in the paper and the Supplementary Materials. Raw data generated during the current study are available from the corresponding authors upon reasonable request.

## Acknowledgements

This work was supported by the National Key Research and Development Program of China (Grant No. 2024YFA1209201 and 2022YFA1604304) and the National Natural Science Foundation of China (Grant No. 12574405 and 92250305). We also thank the State Key Laboratory of Regional Fiber Optic Communication Networks and Novel Optical Communication Systems for nanofabrication facility support.

## Author Contributions

G.Z and X.M contributed equally to this work. G.L and G.Z conceived the idea. G.Z and X.M carried out investigations and numerical simulations. G.Z, Z.C, H.L, X.M and S.H helped with the setup construction. G.Z finished the device fabrication. G.Z and X.M collaborated to complete the measurements. And all authors listed contributed to the manuscript writing, reviewing, and editing.

## Conflict of interest

The authors declare no competing interests.